\documentclass[12pt]{revtex4}
\usepackage{graphicx}

\font\msytwww=msbm5 scaled\magstep1

\let\a=\alpha      \let\e=\varepsilon
     \let\th=\theta  
\let\m=\mu                 
     \let\c=\chi
   
 \let\D=\Delta

\def\EE{{\cal E}} \def\VV{{\cal V}}

\def\hhh{{\bf h}}

\def\rrr{\hbox{\msytwww R}}

\def\hhh={\V h}

\def\\{\hfill\break} \let\==\equiv

\let\io=\infty 

\let\0=\noindent

\def\tende#1{\,\vtop{\ialign{##\crcr\rightarrowfill\crcr
 \noalign{\kern-1pt\nointerlineskip} \hskip3.pt${\scriptstyle
 #1}$\hskip3.pt\crcr}}\,}
\def\circage{\lower2pt\hbox{$\,\buildrel > \over {\scriptstyle \sim}\,$}}
\def\otto{\,{\kern-1.truept\leftarrow\kern-5.truept\to\kern-1.truept}\,}

\def\T#1{{#1_{\kern-3pt\lower7pt\hbox{$\widetilde{}$}}\kern3pt}}
\def\VVV#1{{\VV #1}_{\kern-3pt
\lower7pt\hbox{$\widetilde{}$}}\kern3pt\,}
\def\W#1{#1_{\kern-3pt\lower7.5pt\hbox{$\widetilde{}$}}\kern2pt\,}

\def\indica{\leaders \hbox to 0.5cm{\hss.\hss}\hfill}
\def\guida{\leaders\hbox to 1em{\hss.\hss}\hfill}

\def\VV#1{{\,\underline#1\,}}

\def\to{\rightarrow}

\def\qed{\hfill\raise1pt\hbox{\vrule height5pt width5pt depth0pt}}

\def\indic{\hbox{\raise-2pt \hbox{\indbf 1}}}

\def\V0{{\bf 0}}

\def\V#1{{\bf#1}}

\def\const{{\rm const}}

\def\be{\begin{equation}}
\def\ee{\end{equation}}
\def\bea{\begin{eqnarray}}\def\eea{\end{eqnarray}}
\def\bean{\begin{eqnarray*}}\def\eean{\end{eqnarray*}}
\def\bfr{\begin{flushright}}\def\efr{\end{flushright}}
\def\bc{\begin{center}}\def\ec{\end{center}}
\def\ba#1{\begin{array}{#1}} \def\ea{\end{array}}
\def\bd{\begin{description}}\def\ed{\end{description}}
\def\bv{\begin{verbatim}}\def\ev{\end{verbatim}}

\def\Halmos{\hfill\vrule height10pt width4pt depth2pt \par\hbox to \hsize{}}

\begin{document}

\title{Modulated phases of a 1D sharp interface model\\ in a magnetic field}
\thanks{
\copyright\, 2009  by the authors. This paper may be
reproduced, in its
entirety, for non-commercial purposes.}

\author{Alessandro Giuliani}
\affiliation{Dipartimento di 
Matematica di Roma Tre, Largo S. Leonardo Murialdo 1, 00146 Roma, Italy.}
\author{Joel L. Lebowitz}
\affiliation{Departments of Mathematics and Physics, Rutgers University,
Piscataway, NJ 08854 USA.}
\author{Elliott H. Lieb}
\affiliation{Departments of Physics and Mathematics, 
Princeton University, Princeton, NJ 08544 USA.}
\vspace{3.truecm}
\begin{abstract} We investigate the ground states of 1D continuum models
having short-range ferromagnetic type interactions and a wide class of 
competing longer-range antiferromagnetic type interactions. 
The model is defined in terms of an energy functional, which can be thought of
as the Hamiltonian of a coarse-grained microscopic system or as a mesoscopic 
free energy functional describing various materials.
We prove that the ground state is simple periodic whatever the prescribed total
magnetization might be. Previous studies of this model of frustrated
systems 
assumed this simple periodicity but, as in many examples in condensed matter 
physics, it is neither obvious nor always true that ground states do not
have a more complicated, or even chaotic structure. 
\end{abstract}

\maketitle

\renewcommand{\thesection}{\arabic{section}}

\section{Introduction}\label{sec1}
\setcounter{equation}{0}
\renewcommand{\theequation}{\ref{sec1}.\arabic{equation}}

In two previous papers \cite{GLL1,GLL3} we considered one-dimensional  
discrete and continuum models of classical spin
systems with short and long range competing interactions. We proved 
that, if the long-range interactions are reflection positive and the short 
range interaction is ultralocal (nearest neighbor in the lattice case), 
then the ground states of the system 
display periodic striped order. The proof was based on 
antiferromagnetic reflections about the nodes of the spin density 
configuration, and used 
the fact that no external magnetic field was imposed or, equivalently, 
that the total magnetization was zero. In this note, we extend the analysis
of \cite{GLL1,GLL3} 
to a continuum sharp interface model in the case of non-zero magnetization.
We find that for a large class of antiferromagnetic long range interactions
with arbitrarily fixed total magnetization, all the ground states are 
{\it simple periodic}, i.e., they consist of a sequence
of blocks of alternate sign of the spin and alternate lengths 
$\ldots,\ell_1,\ell_2,\ell_1,\ell_2,
\ldots$, so that the magnetization per unit length, which is specified, 
is $m=(\ell_1-\ell_2)/(\ell_1+\ell_2)$. Recently, Nielsen, Bhatt and Huse
\cite{Huse} studied the dependence
of the period $\ell_1+\ell_2$ on the surface tension in such a 1D sharp 
interface 
model with power law interactions, under the assumption (supported by numerical
evidence) that all the ground states of the system are simple periodic. 
One of our goals here is to prove that their restriction 
to simple periodicity is justified. 

If we give up the continuum nature of the model then,
in general, the simple periodic states are not expected to be the 
ground states of the system. Indeed, for a discrete Ising model with 
long-ranged antiferromagnetic convex interactions, the ground states 
display a complex structure as a function of the prescribed magnetization. 
See \cite{BB, H, PU}.

Simple periodicity cannot, therefore, be taken for granted, and since 
the numerical tests commonly investigate only the local stability of 
not-too-complex periodic structures, it is desirable to have a rigorous 
proof of simple periodicity. 
In this paper we provide such a proof for reflection positive potentials
(including the power-law potentials considered in 
\cite{Huse}) and for perturbations of reflection positive potentials.
Indeed the number of physical models for which periodicity can be 
rigorously proved is very small \cite{M,AM,CO}, and our methods here might lead to other 
useful examples. This is of particular interest in 2D, where 
mesoscopic free energy functionals of the type we consider here
have been proposed as models for micromagnets \cite{Br,DKMO,GD82}, 
diblock copolymers \cite{HS,L80,OK},
anisotropic electron gases \cite{SK04,SK06}, 
polyelectrolytes \cite{BE88}, charge-density waves in 
layered transition metals \cite{Mc75} and superconducting films 
\cite{EK93}. In all these systems, existence of simple periodic
ground states has been argued heuristically 
\cite{Br,BL,EK93,GD82,GP,HS,L80,N01,SK04,SK06}, but there are 
at present only few rigorous results \cite{DKMO,C,ACO,GLL2,Mur}.

The paper is organized as follows. In Section \ref{sec1a} we define the model, 
state the main results in the form of two theorems, and discuss
their significance.
In Section \ref{sec2} we prove the first theorem, for the case
of reflection positive interactions. The proof combines ideas from our previous
papers and from \cite{M,AM,CO}.
In Section \ref{sec3} we prove stability of our results, namely that 
small perturbations of reflection positive interactions 
do not affect the simple periodicity of the ground state. 
In Section \ref{sec4} we discuss the ground state properties of the system at 
small $J$.
In Appendix \ref{appA} we prove some non-degeneracy properties of the 
minimizers, used in the proof of Theorem 1.

\section{Main results}\label{sec1a}
\setcounter{equation}{0}
\renewcommand{\theequation}{\ref{sec1}.\arabic{equation}}

Given $L>0$, we consider the following energy functional:
\bea&& \EE(u)
=\frac{J}2\int_0^L dx\, |u'|+\frac12\int_0^Ldx \int_{-\io}^{+\io} dy \,u(x)\,
v(x-y)\,\widetilde u(y)\;,\label{1.1}\eea
where $J>0$, $v$ is a positive 
potential, and $u$ is a function defined for $0\le x\le L$
that assumes the values $\pm1$, representing the configurations of our 1D 
magnetic system, and $u'$ is its derivative. For any function $u$ with values 
$\pm1$, $\int_0^L dx\, |u'|$ is simply twice the number of times $u(x)$ 
jumps from $+1$ to $-1$ or from $-1$ to $+1$. 

The function $\widetilde u$, in (\ref{1.1}), is the (Neumann) 
extension of $u$ over the whole real axis and is defined as follows. 
Given a function $w$ defined in an interval $I=[a,b]$, 
its Neumann 
extension $\widetilde w$ is obtained from $w$ by iteratively 
reflecting it about the endpoints $a,b$ of $I$ and about their images,
see Fig. 1. 

\begin{figure}[ht]
\hspace{1 cm}
\includegraphics[height=5.2cm]{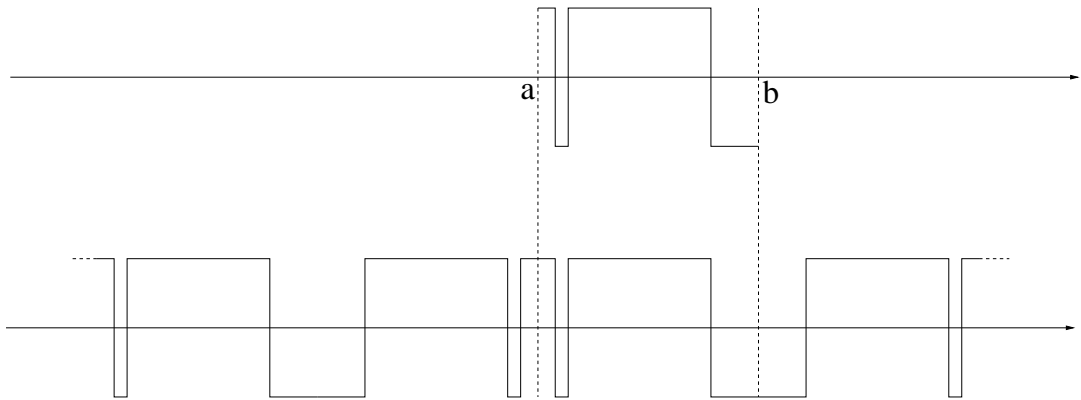}
\caption{A function defined in the interval $[a,b]$ (upper part)
and its Neumann extension (lower part).}
\end{figure}

We will also assume that $u$ satisfies the magnetization constraint:
\be \frac1L\int_0^L\,dx\,u(x)=m\;,\qquad 0\le m<1\;.\label{1.3}\ee 

In the following, we shall require that the potential $v$ 
satisfies some positivity properties. More precisely, we shall consider:
\begin{enumerate}
\item Reflection positive potentials, i.e., 
\be v(x)=\int_0^\io d\a\,\m(\a)\,e^{-\a|x|}\;,\label{1.2}\ee
with $\m$ a positive density such that $v$ 
is integrable, i.e., $\int_0^\io d\a\,\m(\a)\,\a^{-1}<\io$;
(\ref{1.2}) is equivalent
to the condition that $v$ is completely monotone, i.e., $(-1)^n\frac{d^n v(x)}
{dx^n}\ge 0$, for all $x>0$, $n\ge 0$ \cite{Ber}; 
\item Finite range perturbations of reflection positive potentials, i.e., 
\be v(x)=v_0(x)+f_\e(x)\;,\label{1.2a}\ee
with $v_0$ as in (\ref{1.2}) and $f_\e$ a finite,
even potential, with range $\e$. 
\end{enumerate}

Our first result is that in the 
case of reflection positive interactions
the minimizers of (\ref{1.1}) are simple periodic, for all $J>0$. 
\\

{\bf Theorem 1 [Simple periodicity].} 
{\it Given an integer $M$ and $0\le m<1$, 
let $u_{M,m}(x)$ be defined for $0\le x\le L/M$ by
\be u_{M,m}(x)=\cases{+1\hskip.6truecm {\rm if}\quad  0\le x\le 
\frac{1+m}{2}\frac{L}{M};
\cr -1\hskip.6truecm {\rm if} \quad \frac{1+m}{2}\frac{L}{M}\le x\le 
\frac{L}{M}\;.}\label{1.2b}\ee
Then all the finite volume minimizers of (\ref{1.1})
with reflection positive potential (\ref{1.2})
are of the form $w^+_M(x)=\widetilde u_{M,m}(x)$
or $w^-_M(x)=\widetilde u_{M,m}(x-\frac{L}{M})$, 
with $M$ fixed by the variational equation
\be \EE(w^\pm_M)=\min_{M'}\EE(w^\pm_{M'})\;,\label{1.4}\ee
where $M'$ is a positive integer.}

The variational equation (\ref{1.4}) has been studied and solved, for some 
explicit choices of $v$, in \cite{Huse}.\\

One might worry about the fact that the resulting picture of a zero temperature
phase diagram consisting of simple periodic ground states crucially 
depends on the choice of a reflection positive, or at least convex, potential. 
Any reflection positive potential is convex and any convex potential 
that goes to zero at infinity has a cusp at $x=0$. A natural question,
therefore, is whether the cusp plays an important 
role or not  in the resulting phenomenon. It is reassuring that we can
prove that the simple periodicity property is stable under small 
perturbations $f_\e$ of the reflection 
positive potential that remove the cusp, as long as $\e$ is 
smaller than the resulting period.
\\

{\bf Theorem 2 [Perturbative stability].} 
{\it Let $v, v_0$ and $f_\e$ be defined as in (\ref{1.2a}) and
let us assume that 
\be \e< \frac{J}{\int_{-\io}^{\io}dx\,( v_0(x)+2 |f_\e(x)|)}\;.\label{eps}\ee
Then
the finite volume minimizers of (\ref{1.1})
with perturbed reflection positive potential (\ref{1.2a})
are functions of the form $w_M^\pm$, with $w_M^\pm$ defined 
as in Theorem 1, and with $M$ fixed by the variational equation
\be \EE(w^\pm_M)=\min_{M'}\EE(w^\pm_{M'})\;.\label{1.4b}\ee
}

Theorem 2 can be interpreted as saying that for any finite $J$ 
the simple periodicity property is stable under small finite-range 
perturbations of the potential. It can also be interpreted 
the other way round: For any given finite-range perturbation of a reflection 
positive potential, the ground state is 
simply periodic if $J$ is large enough. In this sense, it suffices that the
tails of the long range interaction are ``reflection positive'', in order for 
the ground state to be simply periodic, at least if $J$ is large enough.
On the contrary, at small values of $J$, the structure of the ground states
may depend critically on the short range properties of the potential, as 
discussed in Section \ref{sec3}, after the proof of Theorem 2.

A similar stability result is valid for {\it lattice models} in zero magnetic 
field. Consider a 1D Ising model
with finite range ferromagnetic interactions and long range antiferromagnetic
reflection positive interactions. 
If the strength $J$ of the nearest neighbor (n.n.) ferromagnetic interaction is
large enough, while the strength of the next to nearest neighbor ferromagnetic 
and long range antiferromagnetic interactions are kept fixed, 
the ground states are 
simple periodic. This extends the results of \cite{GLL1}, where simple
periodicity was proved only for the case of n.n. ferromagnetic interactions.
The proof 
of this claim goes along the same lines as the proof of Theorem 2 and we will 
not belabor its details here.

\section{Proof of Theorem 1}\label{sec2}
\setcounter{equation}{0}
\renewcommand{\theequation}{\ref{sec2}.\arabic{equation}}

Let us first fix an integer $M$ and let us temporarily restrict ourselves to 
functions with exactly $M$ jumps in $[0,L]$. Let us rewrite the energy 
of such functions in the form:
\be \EE(u)=JM+\frac12\int_0^\io d\a\,\m(\a)\, E_\a(u)\;,\qquad E_\a(u)=
\int_0^Ldx \,u(x) W_{\a,u}(x)\;,\label{2.1}\ee
where 
\be W_{\a,u}(x)=\int_{-\io}^{+\io} 
dy\, e^{-\a|x-y|}\widetilde u(y)\label{2.2}\ee
is the potential at point $x$ associated to the exponential 
interaction $e^{-\a|x-y|}$. A short calculation shows that $W_{\a,u}$ satisfies
the linear second order equation
\be W_{\a,u}''(x)-\a^2W_{\a,u}(x)=-\a u(x)\;.\label{2.3}\ee
For a given $M$ and $m$ exactly one simple periodic function exists (up to 
translations). We are going to prove that for each $\a>0$, $E_\a(u)$ is 
minimized by this simple periodic function and, therefore, $\EE(u)$
is also minimized by this function. 

Let us now fix $\a$ and let $w$ be a minimizer of 
$E_\a(u)$ in the space of functions with exactly
$M$ jumps. We can assume, without loss of generality, that $w(0)=+1$.
In this case, $w$ is uniquely determined by the sequence of its jump 
points $0\le z_1\le z_2\le \cdots\le z_M\le L$, see Fig.2; 
these jump points have to satisfy a constraint induced by (\ref{1.3}): 
\be z_1-(z_2-z_1)+\cdots+(-1)^{M-1}(z_M-z_{M-1})+(-1)^{M}(L-z_M)=Lm\;.
\label{2.4}\ee

\begin{figure}[ht]
\hspace{1 cm}
\includegraphics[height=2.7cm]{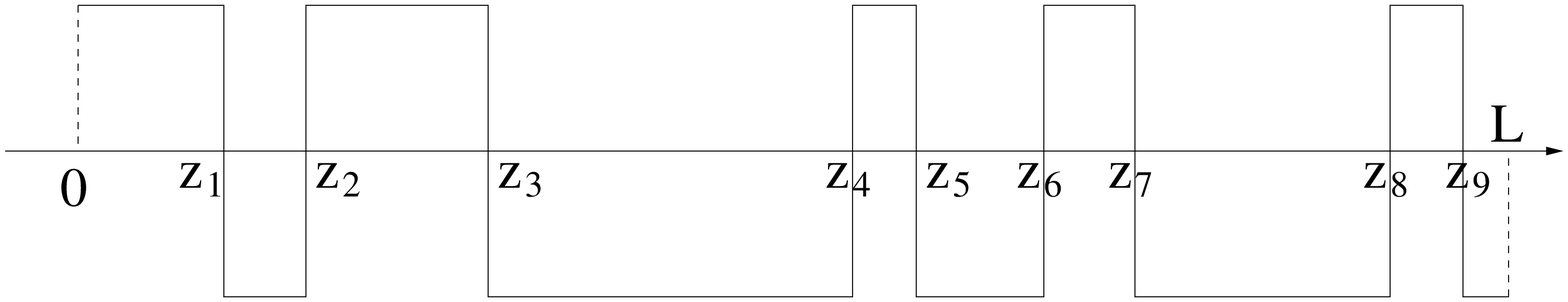}
\caption{A putative minimizer $w$ of $E_\a(u)$ in the subspace
of functions with $M=9$ jumps, and its sequence of non-degenrate jump points.}
\end{figure}

The existence of a minimizer for fixed $\a$ and fixed number of jumps
is proved in Appendix \ref{appA}, where it is shown in particular that
any such minimizer has a non degenerate sequence of jump points, i.e., 
$0<z_1<z_2<\cdots<z_M<L$, and that the potential at the jump points is 
constant, i.e., $W_{\a,w}(z_i)$ is independent of $i$.
As discussed in Appendix \ref{appA}, the potential $W_{\a,w}$ 
is strictly convex in the intervals where $w$ is negative and concave 
in the intervals where $w$ is positive. Therefore, 
$W_{\a,w}$
has exactly one zero derivative point in each interval 
$(z_i,z_{i+1})$, $i=1,\ldots,M-1$; let us denote it by $x_i$, 
$x_i\in(z_i,z_{i+1})$. 
We also define $x_0=0$ and $x_M=L$; note that, by the Neumann's boundary 
conditions imposed on the big box $[0,L]$, we also have that
$W_{\a,w}'(x_0)=W_{\a,w}'(x_M)=0$.

The ordered (and non degenerate) sequence of points 
$x_i$, $i=0,\ldots,M$, induces a partition 
of $[0,L]$ in intervals $I_i=[x_i,x_{i+1}]$ characterized by the fact that 
$W_{\a,w}'(x_i)=0$. Now, the first key remark, due to M\"uller and to Chen and 
Oshita \cite{M,CO}, is that, 
for every $x\in I_i$, $W_{\a,w}(x)=W_{\a,w_{i}}(x)$, where 
$w_{i}=\widetilde w_{I_i}$, with $w_{I_i}$ the restriction of $w$ to $I_i$.
In other words the claim is that, if we restrict to intervals whose endpoints
are zero derivative of the potential, then the potential inside such an 
interval is the same as one would get by repeatedly reflecting $w_{I_i}$ about 
the endpoints of $I_i$ and about their images under reflections. 
The reason is very simple: both $W_{\a,w}(x)$ and 
$W_{\a,w_{i}}(x)$ satisfy the same equation (\ref{2.3}) in the same 
interval, with $W'=0$ boundary consitions at $x_i$ and $x_{i+1}$.
The solution of the linear equation (\ref{2.3})
with these boundary conditions is unique, which means that the
two potentials must be the same on $I_i$. Therefore,
\be\int_0^Lw(x)W_{\a,w}(x)=\sum_{i=0}^{M-1}\int_{x_i}^{x_{i+1}}
w_i(x) W_{\a,w_{i}}(x)\;.\label{2.4a}\ee

\begin{figure}[ht]
\hspace{1 cm}
\includegraphics[height=5.4cm]{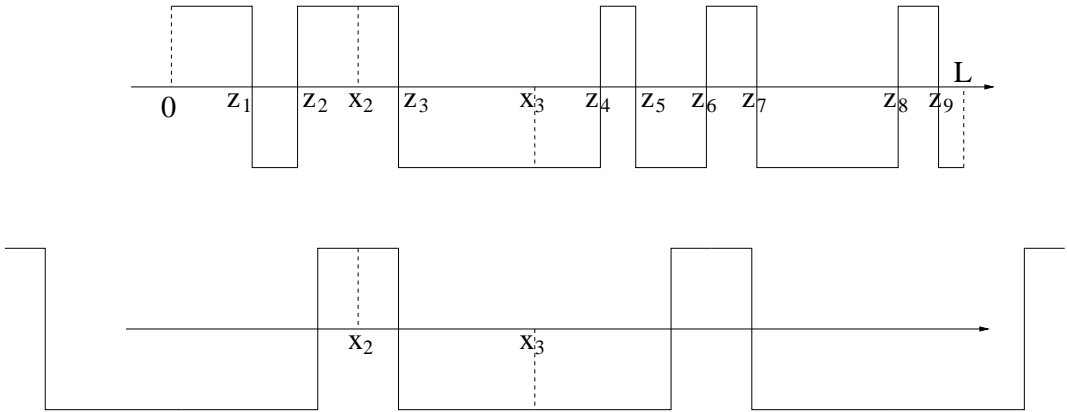}
\caption{A putative minimizer $w$ with $M=9$ jumps (upper part).
If $x_2$ and $x_3$ are zero derivative points of $w$, then the potential 
generated by $w$ and by $\widetilde w_3$ (lower part)
inside the interval $[x_2,x_3]$ are the same.}
\end{figure}

On the other hand, denoting by $p_i,q_i$ the lengths of the positive and 
negative parts of $w_i$ on $I_i$, respectively, a computation
shows that 
\be \a^2\int_{x_i}^{x_{i+1}}
w_i(x) W_{\a,w_{i}}(x)=2p_i\a+2q_i\a-4\frac{
\sinh(\a p_i)\,\sinh(\a q_i)}{\sinh(\a p_i+\a q_i)}\= f(\a p_i,\a q_i)
\;.\label{2.5}\ee
It is straightforward to check that $f$ is a jointly strictly convex 
function of the variables $(p,q)$, that is, the second derivative 
matrix (the Hessian) of $f(x,y)$, which is
\be H(f)(x,y)=\frac{8}{[\sinh(x+y)]^3}\pmatrix{ (\sinh y)^2\cosh(x+y)&
-\sinh x\cdot\sinh y
\cr -\sinh x\cdot \sinh y &(\sinh x)^2 \cosh(x+y)}\;,\label{2.6}\ee
is positive definite for all $x,y>0$. The convexity implies that the minimum 
energy occurs when all the $p_i$ and $q_i$ are the same, 
given the constraint on their sums. Thus, the potential 
energy at fixed $\a$ of a minimizer $\phi$ in the subspace of functions
with $M$ jumps satisfies:
\bea \a^2\int_0^Lw(x)W_{\a,w}(x)=\sum_{i=0}^{M-1}f(\a p_i,\a q_i)
&\ge& Mf(\a\frac{\sum_i p_i}{M}, \a\frac{\sum_i q_i}{M})=\label{2.7}\\
&=&M f(\a\frac{L}{M}\frac{1+m}2, \a\frac{L}{M}\frac{1-m}{2})\;.\nonumber\eea
In the last equality we used the mass constraint (\ref{1.3}). Note that 
the inequality in (\ref{2.7}) is strict unless the values of $(p_i,q_i)$ 
are independent of $i$. Now, the r.h.s. of (\ref{2.7}) is nothing else but 
$\int_0^L dx\int_{\rrr} dy\,w^\pm_M(x)
\,e^{-\a(x-y)} \widetilde w^\pm_{M}(y)$, with $w^\pm_M$ defined 
as in Theorem 1. This shows that the only two minimizers of
$E_\a(u)$ on the subspace of functions with $M$ 
jumps are precisely the $w_M^\pm$ defined in Theorem 1. Quite remarkably,
these minimizers are independent of $\a$: this is the second key remark. 
Therefore, averaging over $\a$ and minimizing over $M$, we get Theorem 1.
Q.E.D.\\

Let us conclude this section by a comment. The proof of Theorem 1 raises
the question of whether there might be non simple periodic ``metastable''
states $w$ in which the potential at the jump points, $W_{w}(z_i)=
\int_0^{\io}d\a\, \m(\a)W_{\a,w}(z_i)$, are all equal. A computation
along the same lines of the proof of Theorem 1 allows one to prove that such
metastable states do not exist when $v(x)=Ce^{-\a_0|x|}$ but we do not know
whether these are possible for more general reflection positive 
(or just convex) potentials.

\section{Proof of Theorem 2}\label{sec3}
\setcounter{equation}{0}
\renewcommand{\theequation}{\ref{sec3}.\arabic{equation}}

Let us fix $J>0$ and let us consider a minimizer $w$ of (\ref{1.1}).
Let $M^*$ be its number of jumps and let $h_0=2z_1$, $h_1=z_2-z_1$,
$\ldots$, $h_{M^*-1}=z_{M^*}-z_{M^*-1}$, $h_{M^*}=2(L-z_{M^*})$ 
be the corresponding block sizes. An important remark is that,
for any fixed $J>0$, under the assumptions of Theorem 2, 
there is an apriori upper bound on the block sizes in the ground state.
In fact, since $w$ 
is an energy minimizer, energy must not decrease if we change sign of 
$w$ in $(z_i,z_{i+1})$, i.e., in the block of size $h_i$. If $\D E$ denotes 
the energy change corresponding to such sign change, we have
\be 0\le \D E\le -2J+4\int_{z_i}^{z_{i+1}}dx\,\int_{z_{i+1}}^{\io}dy\,
(v_0(x-y)+|f_\e(x-y)|)\;.\label{3.1}\ee
Since $f_\e$ has range $\e$, we see that the r.h.s. of (\ref{3.1})
is bounded above by $-2J+2h_i\int_{-\io}^{+\io}dx\,|v_0(x)|+4\e\int_{-\io}^{
+\io}dx\,|f_\e(x)|$, which 
implies 
\be h_i\ge \frac{J-2\e\int_{-\io}^{+\io}dx\,|f_\e(x)|}{\int_{-\io}^{+\io}dx\,
|v_0(x)|}\=h^*\;,\label{3.1z}\ee
with $h^*>0$, by the assumptions of Theorem 2. It then must be true that 
$\frac{1-|m|}2\frac{L}{M^*}\ge h^*$. 

If, as assumed in Theorem 2, the range $\e$ of the perturbation $f_\e$ 
is strictly smaller than $h^*$, then the contribution to 
the ground state energy coming from $f_\e$ is essentially trivial and
is given by:
\be \frac12\int_0^Ldx\int_{\rrr}dy\, w(x) f_\e(x-y)\widetilde w(y)=
L\int_{\rrr}dx f_\e(x)-2M\int_{0}^{\e}dy\int_{-\e}^0dx\, f_\e(y-x)\;.
\label{3.2}\ee
Therefore, defining $J_0=\frac12\int_{0}^{\e}dy\int_{-\e}^0dx\, f_\e(y-x)$,
we can write, 
\be \EE(w)=L\int f_\e+(J-J_0)M+\frac12\int_0^{\io}d\a\,\m(\a)\,
w(x)W_{\a,w}(x)\;.\label{3.3}\ee
Proceeding as in Section \ref{sec2}, and using the fact that 
$\frac{1-|m|}2\frac{L}{M^*}\ge h^*>\e$, we find that the r.h.s. of (\ref{3.3})
is bounded from below by $\EE(w_{M^*}^\pm)$, 
as desired. As in the proof in Section
\ref{sec2}, the bound below is strict, unless $w=w^\pm_{M^*}$. This concludes 
the proof of Theorem 2. Q.E.D.

\section{Discussion}\label{sec4}
\setcounter{equation}{0}
\renewcommand{\theequation}{\ref{sec4}.\arabic{equation}}

Let us fix a perturbation $f_\e$. By Theorem 2, we know that 
for large enough $J$, the ground states are simply periodic. It is 
natural to ask what happens for smaller values of $J$. We claim that in this 
case the nature of the ground state critically depends on the short range 
properties of the potential and, more precisely, it depends on whether $v$ is 
of positive type (i.e., its Fourier transform $\hat v(k)\ge 0$) 
or not. Before we enter a discussion
of this claim, let us remark that even if $f_\e$ is arbitrarily small,
with an arbitrarily small range, the resulting potential $v=v_0+f_\e$ 
can be of either type, depending on the specific properties
of $f_\e$. E.g., let $v_0(x)=e^{-|x|}$, $g_\e$ a positive compactly supported 
even function of range $\e$ and $A^{-1}=\int_{-\io}^{\io}
dx\, \cosh x \,g_\e(x)$, then the potential $w$, given by the convolution 
of $Av_0$ and $g_\e$, $w=A v_0*g_\e$, is continuous, 
equal to $e^{-|x|}$ for $|x|>\e$ and equal to $e^{-|x|}+O(\e^2)$ if $|x|\le\e$.
Moreover, its Fourier transform has the same sign as that of $\hat g_\e$, 
which might or might not be positive. For example, the triangle function 
$g_\e(x)=\max\{0,\e-|x|\}$ has $\hat g_\e\ge 0$, while the square function
$g_\e(x)=\th(\e-|x|)$ is not of positive type. 

We expect that the nature of the ground state at small 
$J$ depends critically on the positivity of $\hat v$. 
To gain some intuition about the
problem let us first look at the case $J=0$ and let us temporarily 
replace the constraint $|u(x)|=1$ by the softer one $|u(x)|\le 1$. 
In this case, if $v$ is of positive type, then the potential term 
$\int_0^Ldx\int_{-\io}^{+\io}
dy\,u(x)v(x-y)\widetilde u(y)$ is happiest when $u$
is constant, i.e., $u\=m$. When $\min\hat v(k)=\hat v(k^*)<0$ then
the potential energy wants $u$ to be 
modulated at the wavelength $k^*$, e.g., $u=m+\const\cos(k^* x)$, 
\cite{GP,BL}. Now, in the presence of the hard constraint 
$|u(x)|=1$, we can get as close as 
we like to this by approaching in a weak limiting sense 
the previous minimizing configurations by a sequence of highly oscillating 
functions $u_i$ that take only the values $\pm1$ but which approximate the 
smooth function $m+\const\cos(k^* x)$. 
Clearly, in the presence of a small positive $J$,
the minimizer will be close to one of these highly oscillating configurations,
with a finite (but possibly very small) oscillation scale. 
Therefore, if $v$ is not of positive type, the minimizer at small $J$ will be
close to a highly oscillating approximation of the aforementioned modulated 
minimizer, {\it which is not simply periodic}. If $v$ is of positive type,
the minimizer at small $J$ will be close to a highly oscillating approximation 
of the constant configuration $u\=m$, and it may very well be that the 
optimal $u$ is simply periodic. We actually conjecture that this 
is the case.

\acknowledgements  We thank S. M\"uller for useful discussions.
The following support is
gratefully acknowledged: U.S. N.S.F. grants PHY-0652854 (E.H.L. and A.G.)
DMR-082120 (J.L. and A.G.). AFOS-FA9550-09 (J.L. and A.G.).

\appendix
\section{Non degeneracy of the minimizers}
\label{appA}
\setcounter{equation}{0}
\renewcommand{\theequation}{\ref{appA}.\arabic{equation}}

In this Appendix we show that, for any $\a>0$, 
the minimizers $w$ of $E_\a(u)=\int_0^Ldx\,u(x)W_{\a,u}(x)$ 
on the subspace of functions with exactly $M$ jumps are associated to 
a non-degenerate sequence of jump points, $z_0\=0<z_1<\cdots<z_M<L\=z_{M+1}$;
in other words, $z_j=z_{j+1}$ does not 
occur. Moreover, $W_{\a,w}(z_i)$, $i=1,\ldots,M$, 
is independent of $i$, as claimed in Section \ref{sec2},
right after (\ref{2.4}). 

Given any $u$ with exactly $M$ jumps (not 
necessarily a minimizer), let us 
identify it with its (possibly degenerate) sequence of jumps. 
This space of ordered sequences is clearly compact, 
so we have at least one minimizing sequence, which can, in principle, be 
degenerate; let us denote it by $0\le z_1\le \cdots\le z_M\le L$.
If this sequence is degenerate, let 
$0<\tilde z_1<\cdots<\tilde z_{M_0}<L$, $M_0<M$, be the non-degenerate
ordered subsequence of $(z_1,\ldots,z_M)$. That is, we throw away the 
degenerate jumps at $z_j=z_{j+1}$. 
In this case, let us denote by $\phi$
the non degenerate function belonging to the subspace of functions with 
$M_0$ jump points, associated to the sequence $\tilde z_1,\ldots,
\tilde z_{M_0}$. Clearly, $\int_0^Ldx\, w(x)W_{\a,w}(x)=
\int_0^Ldx\, \phi(x)W_{\a,\phi}(x)$ and $\phi$
is a minimizer of $E_\a$ in the subspace of functions with 
$M_0$ jumps. With some abuse of notation, we shall denote the energy 
of this non-degenerate configuration, as a function of the position
of its jump points, by $E_\a(\tilde z_1,\ldots,\tilde z_M)$.
By minimality, $\partial_\e E_\a(\tilde z_1,\ldots,\tilde z_i+\e,\tilde z_{i+1}
+\e,\ldots,\tilde z_M)\big|_{\e=0}=0$, which implies that 
$W_{\a,\phi}(z_i)$ is independent of $i$, with $i=1,\ldots,M_0$.  

Now, the potential $W_{\a,\phi}$ is concave
in the intervals where $\phi$ is positive, and convex in the intervals
where $\phi$ is negative, as we shall now prove. 
Assume that $\tilde z_i<x<\tilde 
z_{i+1}$ is such that $\phi(x)=+1$; in this case, rewriting 
$\phi(x)=-1+2\c_{\phi}(x)$, with $\c_{\phi}$ the 
characteristic function of the region where $\widetilde\phi$ is positive,
we have that 
$W_{\a,\phi}(x)=-2\a^{-1}+2\int_{\rrr} 
dy\,e^{-\a|x-y|}\c_{\phi}(x)$,
from which it is apparent that $W_{\a,\phi}(x)$ is convex, being the 
superposition of strictly convex functions. A similar proof applies to 
the case where $x$ is such that $\phi(x)=-1$. As a consequence, 
there is exactly one strict internal maximum of the potential in every interval
where the minimizer is positive, and exactly one strict internal minimum 
in every interval where the minimizer is negative. Therefore, 
we can always decrease the total potential energy by 
adding $M-M_0$ non-degenerate jumps, sufficiently close to each other
and sufficiently close to, say, the left boundary of the big box $[0,L]$;
this contradicts the assumption that $w$ is a minimizer in the subspace of
configurations with $M$ jumps, and proves the claim.

\end{document}